\documentclass[11pt]{article}
\pdfoutput=1

\usepackage{contour}
\usepackage{xcolor}
\usepackage{float}
\usepackage{mathrsfs}
\usepackage{amsmath}
\usepackage{amssymb}

\contourlength{0.05em}

\usepackage{jheppub}
\usepackage{color}
\usepackage{graphicx}
\usepackage{wrapfig,enumerate,slashed}

\usepackage{footmisc}
\usepackage{amsmath}
\usepackage{wasysym} 
\usepackage{graphicx}
\usepackage{subcaption}
\usepackage{color}
\usepackage{comment}
\usepackage{appendix}
\usepackage{slashed}

\newcommand{\be}{\begin{equation}}
\newcommand{\ee}{\end{equation}}
\newcommand{\beq}{\begin{equation}}
\newcommand{\eeq}{\end{equation}}

\newcommand{\bee}{\begin{eqnarray}}
\newcommand{\eee}{\end{eqnarray}}
\newcommand{\beeq}{\begin{equation}}
\newcommand{\eeeq}{\end{equation}}

\DeclareMathOperator{\Tr}{Tr}

\usepackage{tikz}
\usepackage{orcidlink}

\makeatletter
\gdef\@fpheader{}
\makeatother

\begin{document}

\title{Continuous-variable photonic quantum extreme learning machines for fast collider-data selection}

\preprintA{IPPP/25/63}

\author[a]{Benedikt Maier\,\orcidlink{0000-0001-5270-7540}\:}
\author[b]{Michael Spannowsky\,\orcidlink{0000-0002-8362-0576}\:}
\author[b]{Simon Williams\,\orcidlink{0000-0001-8540-0780}\:}

\emailAdd{simon.j.williams@durham.ac.uk}

\affiliation[a]{\vspace{0.1cm} Blackett Laboratory, Imperial College, Prince Consort Road, London, SW7 2AZ, United Kingdom}
\affiliation[b]{\vspace{0.1cm} Institute for Particle Physics Phenomenology, Durham University, Durham DH1 3LE, UK}
	
\abstract{We study continuous-variable photonic quantum extreme learning machines as fast, low-overhead front-ends for collider data processing. Data is encoded in photonic modes through quadrature displacements and propagated through a fixed-time Gaussian quantum substrate. The final readout occurs through Gaussian-compatible measurements to produce a high-dimensional random feature map. Only a linear classifier is trained, using a single logistic regression, so retraining is fast, and the optical path and detector response set the analytical and inference latency. We evaluate this architecture on two representative classification tasks, top-jet tagging and Higgs-boson identification, with parameter-matched multi-layer perceptron (MLP) baselines. Using standard public datasets and identical train, validation, and test splits, the photonic Quantum Extreme Learning Machine (QELM) outperforms an MLP with two hidden units for all considered training sizes, and matches or exceeds an MLP with ten hidden units at large sample sizes, while training only the linear readout. These results indicate that Gaussian photonic extreme-learning machines can provide compact and expressive random features at fixed latency. The combination of deterministic timing, rapid retraining, low optical power, and room temperature operation makes photonic QELMs a credible building block for online data selection and even first-stage trigger integration at future collider experiments.
}

\maketitle


\section{Introduction}

In particular, continuous-variable quantum computing (CVQC), which encodes information in the quadrature variables of quantum oscillators called \textit{qumodes}, offers a natural route to fast, highly parallel pre-processing. Information is carried and transformed optically, enabling nanosecond-scale operation governed by device transit times and optical control rather than firmware compile cycles. The underlying theory and toolbox for quantum information and computation using continuous variables (CV) are mature, including both Gaussian operations, such as displacement, squeezing, linear optics, and non-Gaussian resources needed for universality~\cite{Lloyd1999,Braunstein2005,Weedbrook2012,Adesso2014,Menicucci2006}. Already, CVQC has seen applications in high-energy physics, including event classification~\cite{Blance:2020ktp} and the simulation of scalar field theories~\cite{Abel:2024kuv, Abel:2025zxb, Abel:2025pxa}. 

\begin{figure}[t!]
    \centering
    \includegraphics[width=\textwidth]{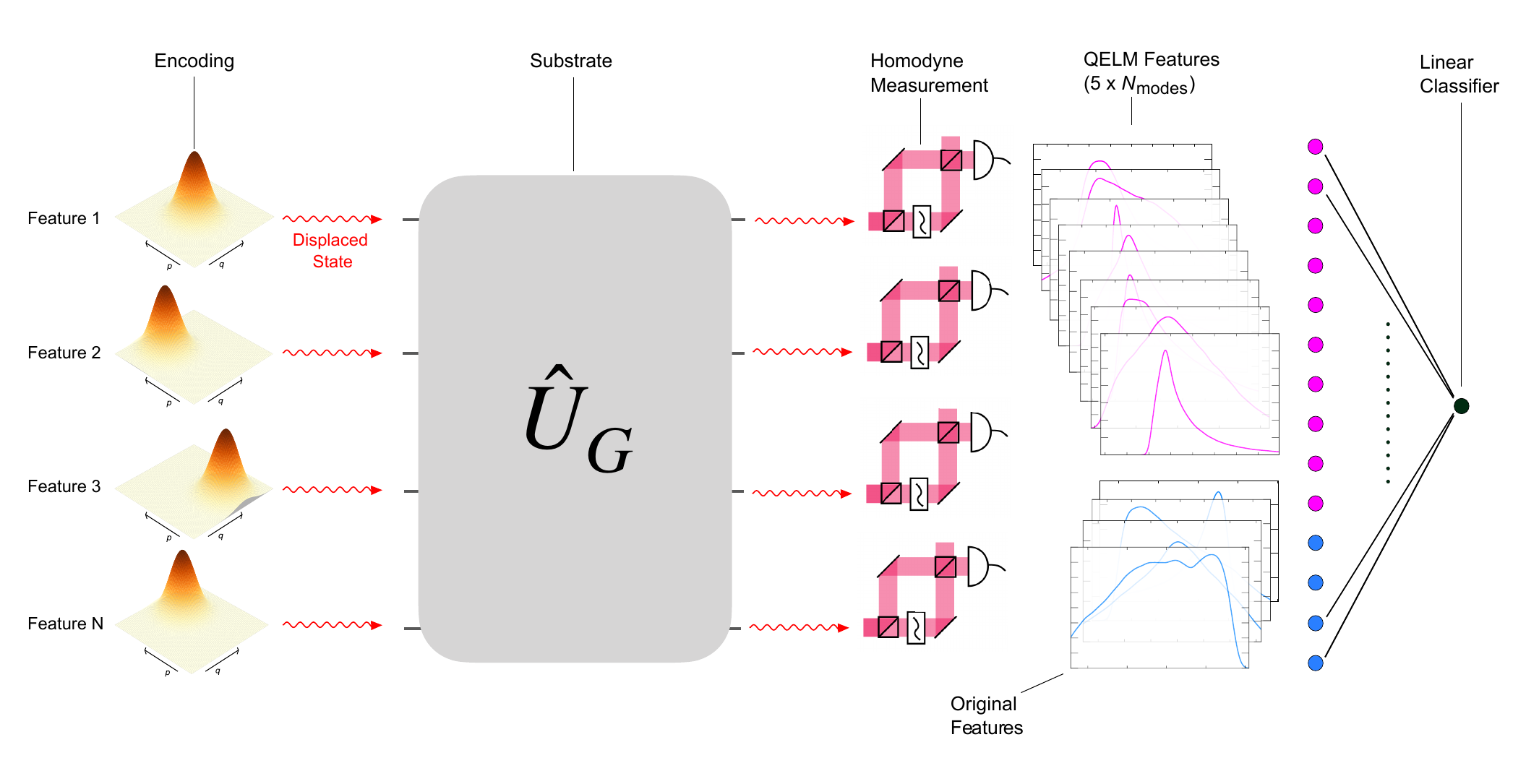}
    \caption{Visualisation of the QELM. The features of the data get encoded onto qumodes via displacement operations. A fixed-time Gaussian quantum substrate $\hat{U}_G$ followed by Gaussian-compatible measurements creates a random, high-dimensional feature map, which is used, together with the original data, in a linear classifier.}
    \label{fig:setup}
\end{figure}

From a machine learning perspective, extreme learning machines (ELMs) \cite{Huang2006, Huang2012, Guo2015} provide a particularly attractive model class for low-latency inference and ultra-fast retraining. ELMs fix a random hidden layer and learn only a linear readout by solving a single regression, thereby eliminating back-propagation and its associated compute and tuning overhead~\cite{Huang2006}. When the random feature map is provided by a physical device, the approach inherits the device’s speed, parallelism, and analog bandwidth. Photonic platforms are especially compelling in this role, with extensive progress in nanophotonic neural primitives and integrated photonic reservoir computing that demonstrate low-power, high-throughput inference compatible with on-edge deployment~\cite{Shen2017,Henderson2020,Vandoorne2014}. 

This connects naturally to the quantum extreme learning machine (QELM) framework \cite{Cindrak2024,Xiong2025}, where a fixed, possibly uncalibrated quantum circuit produces a high-dimensional quantum random feature map that is read out and linearly trained. In this sense, QELMs instantiate the broader idea of quantum feature maps underpinning quantum kernel methods~\cite{PhysRevLett.122.040504,Havlicek:2018nqz}. Theory has clarified both potential and limits of QELMs, including concise descriptions via effective measurements, expressivity analyses, and guidance on circuit design to avoid concentration effects~\cite{Mujal2021,Innocenti2023,Xiong2025}. First applications already demonstrate state estimation and entanglement witnessing with photonic implementations, highlighting the practicality of the train-by-regression paradigm on near-term devices \cite{Vetrano2025,Zia2025,Assil2025}.  

Within high-energy physics, fast artificial intelligence (AI) on specialised hardware has matured from concept to deployed tools for triggers and real-time reconstruction, with the \texttt{hls4ml} dataset~\cite{pierini_2020_3602260} providing a pathway from trained models to implementations on field programmable gate arrays (FPGA) and application-specific integrated circuits (ASIC) under strict latency budgets~\cite{Duarte2018,Aarrestad2021}. In parallel, photonic computing has demonstrated integrated reservoirs and coherent optical neural networks with competitive accuracy and orders-of-magnitude speed and energy benefits in the lab~\cite{Vandoorne2014,Shen2017}. On the quantum side, QELMs and quantum reservoir computing synthesise these strands by using the inherent dynamics and large Hilbert spaces of quantum circuits to implement powerful random feature maps with minimal training, a direction surveyed in~\cite{Mujal2021} and formalised in~\cite{Innocenti2023,Xiong2025}. Recent photonic QELM experiments underscore the feasibility for quantum state tasks~\cite{Vetrano2025,Zia2025,Assil2025}. Our work brings this emerging paradigm into the collider context, proposing and evaluating a continuous-variable photonic QELM as an ultra-fast, reconfigurable front-end for pattern recognition at the detector edge.

We present a continuous-variable quantum extreme learning machine (CV-QELM) whose quantum circuit is a fixed-time Gaussian photonic reservoir, called a quantum substrate. Inputs are encoded as quadrature displacements across $M$ optical modes, qumodes, which propagate through randomly initialised layers of two-mode Gaussian operations that generate entanglement and mode mixing. The qumodes are then measured using either photon-number-resolving (PNR) or homodyne detection to produce a random feature vector, $\Phi(\mathbf{x})$, of dimension $R$, such that the quantum substrate generates a high-dimensional random feature map,
\begin{equation}
\mathbf{x}\in\mathbb{R}^{d}\;\longmapsto\; \Phi(\mathbf{x})\in\mathbb{R}^{R}~.
\end{equation}
Training on the linear readout is then performed via either logistic regression or ridge regression, providing analytic, closed-form updates that make retraining extremely fast. A schematic of the setup is shown in Figure~\ref{fig:setup}. This design matches the requirements of front-end data processing, where latency is set by the optical path and detector response, power consumption is dominated by low optical powers and photodiodes, and reconfiguration is achieved by adjusting homodyne phases and pump settings rather than recompiling firmware. 

The approach is motivated by two observations. First, ELM-style training is attractive under strict latency and retraining constraints, since only the readout is learned, allowing rapid adaptation to changing detector conditions and domain shifts in the data stream. Second, Gaussian photonic circuits supply a physically native, massively parallel random feature map with large effective bandwidth. Although classically simulable, such circuits act as analog random projections that approximate families of smooth kernels while operating at optical line rates. The photonic implementation therefore targets speed, stability, and parallelism rather than quantum advantage, which aligns with on-detector use cases.

Relative to the state of the art, this work focuses on establishing a controlled and reproducible evaluation of CV-QELMs for collider applications. The photonic substrate is designed to operate at fixed optical depth, ensuring that latency is determined solely by path length and detector response, consistent with Level-1 trigger constraints. The random feature map is drawn from a well-defined ensemble of Gaussian photonic circuits, providing a controlled and reproducible framework for assessing quantum random feature models. To isolate the contribution of these photonic features from pure model capacity, we compare against parameter-matched MLPs trained on identical data splits and quantify accuracy as a function of training sample size. This framework provides a clear, reproducible baseline for future work that will extend to systematic sweeps over substrate parameters, direct latency measurement, and hardware implementation studies.

Section~\ref{sec:qelm} reviews ELMs and QELMs and sets our notation. Section~\ref{sec:cvqelm} details the CV photonic substrate, data embedding, and readout. Section~\ref{sec:results} presents results for top-jet tagging and Higgs event identification, including accuracy versus sample size and parameter-matched comparisons to classical baselines. Section~\ref{sec:summary} summarizes findings and outlines next steps.


\section{Quantum Extreme Learning Machines}
\label{sec:qelm}

We begin with the classical extreme learning machine (ELM) formalism~\cite{Huang2006, Huang2012, Guo2015} before transitioning to its quantum analogue. Let \(\{(\mathbf{x}_i, y_i)\}_{i=1}^N\subset\mathbb{R}^d\times\mathbb{R}\) be a training set, or more generally \(y_i\in\mathbb{R}^C\) for classification into \(C\) classes. An ELM is a single‐hidden‐layer feedforward model whose output takes the form
\begin{equation}
f(\mathbf{x}) = \sum_{j=1}^L \beta_j\, h_j(\mathbf{a}_j,\mathbf{x}) \;=\; \boldsymbol{\beta}^\top \mathbf{h}(\mathbf{x})~,
\end{equation}
where \(\mathbf{h}(\mathbf{x}) = (h_1(\mathbf{a}_1,\mathbf{x}), \dots, h_L(\mathbf{a}_L,\mathbf{x}))^\top\) is a vector of nonlinearly transformed features. In the ELM paradigm, the hidden mapping parameters \(\{\mathbf{a}_j\}_{j=1}^L\) are chosen randomly, e.g.\ from a Gaussian or uniform distribution, and held fixed. Only the output weights \(\boldsymbol\beta\) are trained. Collecting the hidden outputs over the training set into the matrix,
\begin{equation}
H_{ij} = h_j(\mathbf{a}_j, \mathbf{x}_i), \quad i = 1,\dots, N,\; j = 1,\dots,L~,
\end{equation}
and putting the target vector(s) in a matrix \(T\in \mathbb{R}^{N\times C}\), the training reduces to solving,
\begin{equation}
H \boldsymbol\beta = T~,
\end{equation}
or its regularised version. A standard choice is ridge regression,
\begin{equation}
\hat{\boldsymbol\beta} = \arg\min_\beta \, \|H \beta - T\|_2^2 + \lambda \|\beta\|_2^2
\;=\; (H^\top H + \lambda I)^{-1} H^\top T~.
\end{equation}

As only $\boldsymbol\beta$ is learned, ELMs avoid iterative gradient descent, backpropagation, learning-rate tuning, and convergence criteria. Under suitable assumptions on the random basis $h_j$, such as bounded activation functions and independent random weight draws, ELMs satisfy universal approximation theorems. They can achieve good generalisation performance for sufficiently large hidden-layer width $L$ \cite{Huang2006,Huang2012,Guo2015,Scardapane2017}. Rigorous analyses have established that, with probability one, randomly generated hidden nodes span dense function spaces in $L_2$ and $C(X)$ norms~\cite{Huang2012,Guo2015}, and that the ELM output converges to the target function as $L\!\to\!\infty$. Moreover, kernel formulations show that ELMs implicitly implement a random feature approximation to certain reproducing-kernel Hilbert spaces, linking their generalisation bounds to those of random-feature models~\cite{Huang2012,Scardapane2017,Rahimi2008}. These results formalise why, despite their conceptual simplicity, ELMs can approach the performance of deeper networks while retaining analytic and hardware efficiency.

The QELM generalises this architecture by replacing the fixed non-linear feature map \(\mathbf{h}(\cdot)\) with a quantum random feature map produced by a quantum substrate or circuit. Concretely, one defines a parametric quantum encoding map
\begin{equation}
\mathbf{x} \mapsto \rho(\mathbf{x}) = U_{\rm enc}(\mathbf{x})\,\rho_0\,U_{\rm enc}^\dagger(\mathbf{x})~,
\end{equation}
evolves under a fixed (random) quantum reservoir unitary \(U_{\rm res}\), and then measures a set of observables \(\{O_k\}_{k=1}^M\). The predicted output is
\begin{equation}
f_{\boldsymbol{\eta}}(\mathbf{x}) = \sum_{k=1}^M \eta_k \,\Tr\bigl[O_k\, U_{\rm res}\,\rho(\mathbf{x})\,U_{\rm res}^\dagger\bigr]
= \boldsymbol{\eta}^\top \mathbf{o}(\mathbf{x})~,
\end{equation}
where the quantum feature vector is \(\mathbf{o}(\mathbf{x}) = ( \Tr[O_1 \rho_{\rm res}(\mathbf{x})], \dots, \Tr[O_M \rho_{\rm res}(\mathbf{x})] )^\top\). The coefficients \(\boldsymbol\eta\in\mathbb{R}^M\) are obtained by a linear solve akin to ridge regression:
\begin{equation}
\hat{\boldsymbol\eta} = (O^\top O + \lambda I)^{-1} O^\top T~,
\end{equation}
where \(O_{ik} = \Tr[O_k \rho_{\rm res}(\mathbf{x}_i)]\). Thus, QELMs preserve the `train-only-the-output-layer' philosophy of ELMs, but now the feature map is determined by quantum dynamics.

One principal advantage of QELMs is the ability of the quantum substrate to implicitly explore extremely large Hilbert spaces, and to embed non-trivial correlations, interference, and, in non-Gaussian or non-linear settings, entanglement structure, which may yield more expressive features than classical random projections for certain tasks. Of course, these advantages bring caveats. Recent work \cite{Xiong2025} develops a Fourier‐expressivity analysis showing that the prediction \(f_{\boldsymbol\eta}\) can be expanded in a Fourier series whose accessible frequencies are determined by the encoding scheme, while the effective number of independent coefficients is bounded by \(\min(M,\lvert\Omega\rvert)\), where \(\Omega\) is the frequency set induced by data encoding. This places an intrinsic ceiling on complexity when \(M\), the number of observables, is small. Moreover, concentration phenomena in large quantum systems, due to randomness, noise, entanglement, or global measurements, can degrade independence of the quantum features, effectively collapsing the model to an input-agnostic constant \cite{Xiong2025,Innocenti2023}. Complementary work uses Krylov‐dimension or spread complexity to bound effective expressivity in QELMs and quantum reservoir computers, suggesting that beyond a certain system depth or randomness, marginal expressivity gains saturate \cite{Cindrak2024}.  

In view of these theoretical constraints, a practical QELM must carefully balance the choice of encoding scheme, the number of observables measured, the depth and mixing of the substrate, and regularisation. In this paper we adopt a continuous variable photonic implementation, embedding classical inputs via quadrature displacements, evolving under a fixed Gaussian circuit, and performing Gaussian compatible measurement techniques to define \(\{O_k\}\). We test the quantum substrate's expressivity and robustness by comparing the CV-QELM performance against parameter-matched MLPs on collider classification tasks.


\section{Continuous-variable quantum extreme learning machines}
\label{sec:cvqelm}

Continuous-variable quantum computing (CVQC) provides a natural substrate for quantum extreme learning machines (QELMS), as it encodes information in the continuous quadratures of quantum oscillators, so-called \textit{qumodes}, whose Hilbert space is infinite-dimensional. Here we focus on a quantum optical platform as an experimentally viable realisation of CVQC, where information is carried by photonic modes and the quadrature operators are defined via the creation and annihilation operators,
\begin{align}
    \hat{x} = \frac{1}{\sqrt{2}}(\hat{a} + \hat{a}^\dagger)~, 
    \qquad
    \hat{p} = \frac{i}{\sqrt{2}}(\hat{a}^\dagger - \hat{a})~,
\end{align}
which obey the canonical commutation relation $[\hat{x}_i,\hat{p}_j]=i\delta_{ij}$ (where here we adopt natural units $\hbar=\omega=m=1$). Gaussian operations, generated by Hamiltonians that are at most quadratic in the quadrature operators $\hat{x}$ and $\hat{p}$, act linearly on the quadrature phase space and include displacement, squeezing, and passive linear optics. Their general unitary form can be written as,
\begin{equation}
    \hat{U}_G = \exp\left( -\frac{i}{2}\,\hat{\boldsymbol{r}}^\top G\,\hat{\boldsymbol{r}} + i \boldsymbol{d}^\top \hat{\boldsymbol{r}} \right)~,
\end{equation}
where $\hat{\boldsymbol{r}}=(\hat{x}_1,\hat{p}_1,\dots,\hat{x}_M,\hat{p}_M)^\top$, $G$ is a real symmetric matrix encoding quadratic couplings, and $\boldsymbol{d}$ a displacement vector. The corresponding symplectic transformation acts on first and second moments as $\boldsymbol{r}\to S\boldsymbol{r}+\boldsymbol{d}$ with $S\in\mathrm{Sp}(2M,\mathbb{R})$ \cite{Braunstein2005,Weedbrook2012,Adesso2014}. 

Single-mode Gaussian gates include displacement $\hat{D}(\alpha)=\exp(\alpha\hat{a}^\dagger-\alpha^*\hat{a})$ and squeeze $\hat{S}(z)=\exp\!\left[\tfrac{1}{2}(z^*\hat{a}^2-z(\hat{a}^\dagger)^2)\right]$ operations, with complex parameters $\alpha=re^{i\phi}$ and $z=r'e^{i\phi'}$. Two-mode Gaussian operations generate entanglement, such as the controlled-addition gate
\begin{align}
    \hat{C}_X(s)=\exp(-i s\,\hat{x}_1\otimes\hat{p}_2)~,
\end{align}
which can be decomposed into a series of squeeze operations and beam splitters, such that
\begin{align}\label{eqn:cxgatedecomp}
    \hat{C}_X(s)=\hat{B}\!\left(\tfrac{\pi}{2}+\theta,0\right)[\hat{S}(r)\!\otimes\!\hat{S}(-r)]\,\hat{B}(\theta,0)~,
\end{align}
where $(\sinh r = -s/2)$ and $(\cos(2\theta) = -\tanh r)$ with the beam splitter defined as $\hat{B}(\theta,\phi)=\exp[\theta(e^{i\phi}\hat{a}_1\hat{a}_2^\dagger - e^{-i\phi}\hat{a}_1^\dagger\hat{a}_2)]$. These gates, along with phase shifters and interferometers, form the building blocks of Gaussian photonic circuits that are experimentally accessible with current integrated optics technology. Non-Gaussian gates, generated by higher-than-quadratic Hamiltonians, such as cubic-phase or Kerr interactions, are required for universal quantum computation~\cite{Lloyd1999,Menicucci2006}, but are experimentally demanding due to weak optical non-linearities. In practice, measurement-induced or hybrid machine-learning-based schemes are often employed to emulate them~\cite{PhysRevA.100.012326,PhysRevA.100.052301,Abel:2024kuv,Abel:2025zxb,Abel:2025pxa}, although these are not necessary for the QELMs considered here.

\begin{figure}[t!]
    \centering
    \includegraphics[scale=1]{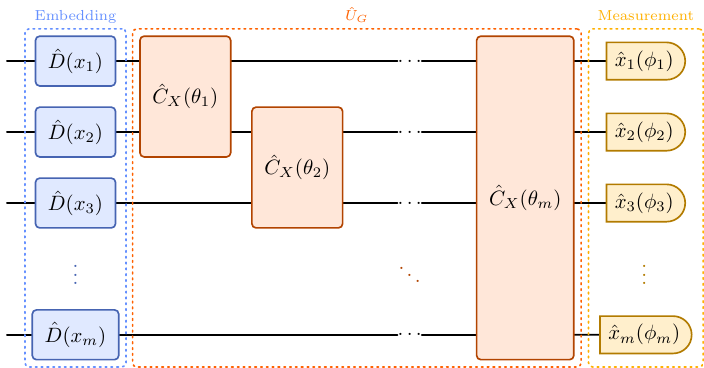}
    \vspace{10pt}
    \caption{Schematic of the photonic substrate for the quantum extreme learning machine (QELM). Classical data are first embedded into optical modes via displacement operations before being propagated through a Gaussian circuit composed of cascading controlled-addition ($\hat{C}_X$) gates. Homodyne measurements are then performed on the output qumodes to generate the feature vector, which serves as input for downstream classical classification.}
    \label{fig:substrate}
\end{figure}

In a continuous-variable QELM, classical data $\mathbf{x}\in\mathbb{R}^d$ are embedded onto the optical device via displacements $U_\text{enc}(\mathbf{x})=\hat{D}(\alpha(\mathbf{x}))$ across $M$ qumodes. The state then evolves under the Gaussian circuit,
\begin{equation}
  \hat{U}_G = \prod_{m=1}^{M}\!\left[\hat{C}_{X_{m , \overline{m + 1}}} (\theta_m)\right]~,  
\end{equation}
comprising a cascade of $C_X$-gates with gate parameters $\theta_m$ which are randomly drawn from experimentally realisable ranges and then fixed. Here, 
\begin{equation}
    \overline{m + 1} = (m+1) \bmod{M}~. 
\end{equation}
denotes the periodicity of the gate structure that provides both mixing and entanglement. We adopt the cascading $C_X$ architecture for three practical reasons: it can be implemented efficiently within Gaussian photonics via the decomposition in Eq.~\eqref{eqn:cxgatedecomp}, it couples all $M$ qumodes with total gate depth proportional to $M$, supporting low-latency operation, and it uses only $M$ random parameters, which keeps the substrate simple to calibrate and reproduce. A schematic of the quantum substrate is shown in Figure~\ref{fig:substrate}. 

For a given input $\mathbf{x}$, the resulting output density operator is
\begin{equation}
    \rho_{\mathrm{out}}(\mathbf{x})=\hat U_G\,\hat U_\text{enc}(\alpha(\mathbf{x}))\,\rho_0\,\hat U_\text{enc}^\dagger(\alpha(\mathbf{x}))\,\hat U_G^\dagger~,
\end{equation}
where $\rho_0 = \vert 0 \rangle\!\langle 0 \vert$ is the initial state of the photonic system where all qumodes are initialised in the vacuum state. 

Finally, the output modes can be measured in two ways. The first method uses PNR measurements on $M$ output modes to determine the occupation of each mode in the photon number basis, yielding discrete outcomes $\mathbf{n}\in\mathbb{N}^M$ sampled from 
\begin{align}
    P(\mathbf{n}\,|\,\mathbf{x})=\Tr\!\left[\rho_{\mathrm{out}}(\mathbf{x})\,\Pi_{\mathbf{n}}\right]~,
\end{align}
where, 
\begin{align}
    \Pi_{\mathbf{n}}=\bigotimes_{m=0}^{M-1} |n_m\rangle\!\langle n_m|~,
\end{align}
are ideal number-resolving measurements. From repeated shots of the circuit, the first and second-order photon-number moments can be obtained as,
\begin{align}\label{eqn:pnrMeas}
    \Phi(\mathbf{x})= \Tr\!\left[\rho_{\mathrm{out}}(\mathbf{x})\,\hat{\mathbf{O}}\right]~,\qquad
    \hat{\mathbf{O}}=\big(\hat n_1,\dots,\hat n_{M},\left(\Delta\hat n_1\right)^{\,2},\dots,\left(\Delta \hat n_{M} \right)^{\,2}\big)^\top~,
\end{align}
where $\hat n_m=\hat a_m^\dagger \hat a_m$ and $\left(\Delta \hat n_m \right)^{\,2} = \hat n_m^{\,2} - \langle \hat n_m \rangle^{\,2}$. The resulting expectation values yield the mean photon numbers, $\bar n_m=\langle \hat n_m\rangle$, and variances, $\mathrm{Var}(\hat n_m)=\langle \hat n_m^{\,2}\rangle-\langle \hat n_m\rangle^2$, for each mode, which together form the classical feature vector $\Phi(\mathbf{x})$, which can then be used downstream for classification.

The second measurement method employs homodyne detection of selected quadratures $\{\hat{x}_m(\phi_m)\}_{m=0}^{M-1}$ on the output modes, yielding continuous-valued outcomes sampled from the probability distribution,
\begin{align}
    P(\mathbf{x}_{\mathrm{hom}}\,|\,\mathbf{x}) = \Tr\!\left[\rho_{\mathrm{out}}(\mathbf{x})\,\Pi_{\mathbf{x}_{\mathrm{hom}}}\right]~, 
\end{align}
where,
\begin{align}
    \Pi_{\mathbf{x}_{\mathrm{hom}}} = \bigotimes_{m=0}^{M-1} \vert x_m(\phi_m)\rangle\!\langle x_m(\phi_m)\vert~, 
\end{align}
denotes the operation corresponding to simultaneous quadrature measurements with phases $\{\phi_m\}$\footnote{The measurement phase determines which quadrature, or which linear combination of quadratures, is measured: \[\hat{x} (\phi) = \cos \phi \,\hat{x} + \sin \phi \,\hat{p}~.\] Note, here we consider measurements of the pure quadratures only, corresponding to $\phi_m=\left\{0,\frac{\pi}{2}\right\}$.}. From these measurements, the first and second-order quadrature moments are obtained as
\begin{align}\label{eqn:homMeas}
    \Phi_\mathrm{hom}(\mathbf{x})=\Tr\!\left[\rho_{\mathrm{out}}(\mathbf{x})\,\hat{\mathbf{O}}_\mathrm{hom}\right]~,\qquad 
    \hat{\mathbf{O}}_\mathrm{hom}
    =\big(&\hat x_0,\dots,\hat x_{M-1},\;
    \hat p_0,\dots,\hat p_{M-1}, \nonumber\\
    &\hat x_0^{2},\dots,\hat x_{M-1}^{2},\;
    \hat p_0^{2},\dots,\hat p_{M-1}^{2}, \nonumber\\
    &\tfrac12\{\hat x_0,\hat p_0\},\dots,\tfrac12\{\hat x_{M-1},\hat p_{M-1}\}\big)^\top~.
\end{align}
This construction captures the local quadrature statistics for each qumode, including both mean values and second-order correlations between conjugate quadratures. The resulting expectation values define the continuous-variable feature vector $\Phi_\mathrm{hom}(\mathbf{x})$, which serves as input to the downstream classification.

Although Gaussian unitaries act linearly on the quadrature phase space of qumodes, the measured observables, shown in Equations~\eqref{eqn:pnrMeas} and \eqref{eqn:homMeas}, depend quadratically on the input displacements, rendering the feature map, $\Phi_\mathrm{hom}(\mathbf{x})$ polynomial in the inputs up to degree two. The expressivity of the CV-QELM therefore arises primarily from the high dimensionality of this polynomial feature map rather than higher-order non-linearities.

Compared to PNR detection, homodyne detection offers a highly efficient and scalable means of extracting continuous-valued features from Gaussian photonic circuits. Its high detection efficiency and access to quadrature correlations make it well suited to near-term implementations of QELMs, where the underlying states remain approximately Gaussian. PNR detection, by contrast, can capture genuinely non-Gaussian statistics at the cost of greater experimental complexity. Homodyne detection also offers an advantage in measurement speed, crucial for the collider applications discussed in this paper. Continuous quadrature measurements can be performed at high bandwidth using standard photodiodes, providing rapid acquisition of statistical moments~\cite{PhysRevA.42.474, PhysRevA.43.330, Weedbrook2012, Huang:15}. In contrast, PNR detection requires slower single-photon counting with limited repetition rates~\cite{Hadfield:23}. Moreover, for Gaussian photonic circuits, homodyne expectation values can be computed efficiently from covariance matrices, making this approach particularly attractive for high-throughput computing. 

Once the data have been propagated through the quantum substrate, classification is performed using a simple classical linear classifier applied to the feature matrix $\Phi_{(\mathrm{hom})}(\mathbf{x})$ obtained from the quantum measurements. Here, we consider two approaches: logistic regression and ridge regression. The former obtains the model parameters $\boldsymbol{\eta}$ by maximising the regularised log-likelihood,
\begin{equation}
\mathcal{L}(\boldsymbol{\eta}) =
\sum_{i=1}^{N}
\Big[
T_i \log \sigma\!\left(\boldsymbol{\eta}^\top \Phi(\mathbf{x}_i)\right)
+ (1 - T_i)\log\!\left(1 - \sigma\!\left(\boldsymbol{\eta}^\top \Phi(\mathbf{x}_i)\right)\right)
\Big]
- \frac{\lambda}{2}\|\boldsymbol{\eta}\|^2~,
\end{equation}
where $N$ is the number of samples, $T_i \in \{0,1\}$ is the target label, $\sigma(z) = 1/(1 + e^{-z})$ is the logistic sigmoid, and $\lambda$ is the regularisation strength. The objective is convex in $\boldsymbol{\eta}$, ensuring a unique global optimum that can be efficiently reached via standard gradient-based solvers. 

In contrast, ridge regression determines the model parameters $\boldsymbol{\eta}$ analytically, avoiding iterative gradient descent and thereby reducing the computational latency of the classification process. This is achieved by minimising the regularised least-squares objective,
\begin{equation}
\mathcal{L}(\boldsymbol{\eta}) =
\sum_{i=1}^{N}
\left(
T^\prime_i - \boldsymbol{\eta}^\top \Phi(\mathbf{x}_i)
\right)^{\!2}
+ \lambda \|\boldsymbol{\eta}\|^2~,
\end{equation}
where $T^\prime i \in \{-1, +1\}$ denotes the class label for the $i$-th training sample and $\lambda > 0$ once again controls the strength of the L2 regularisation. This objective defines a convex quadratic optimisation problem that admits a closed-form solution,
\begin{equation}
\boldsymbol{\eta} = \left(\Phi^\top \Phi + \lambda I\right)^{-1} \Phi^\top \mathbf{T},
\end{equation}
found directly via linear algebra rather than iterative gradient descent. The predicted class for a new input $\mathbf{x}$ is then given by
\begin{equation}
\hat{T} = \mathrm{sign}\!\left(\boldsymbol{\eta}^\top \Phi(\mathbf{x})\right).
\end{equation}
In both formulations, the Gaussian photonic circuit provides a fixed, random non-linear embedding of the data, while the regression models constitute the learnable output layer.

The use of Gaussian circuits for QELMs offers several advantages. They are experimentally stable, with operations that can be easily reconfigured via phase and pump control. Moreover, they operate on continuous-variable optical modes, enabling natural parallel encoding of real-valued input features. Even though purely Gaussian dynamics are classically simulable, their random projections act as efficient high-dimensional analogue feature maps that approximate smooth kernel families~\cite{Killoran2019,Henderson2020,Shen2017}. For real-time or on-detector use cases, this physical implementation emphasises determinism and latency over quantum advantage: the optical computation proceeds at the speed of light, and retraining requires simple linear regression, providing several orders of magnitude faster adaptation than iterative gradient-based neural networks. Compared to digital accelerators, the photonic substrate introduces negligible thermal load and can operate at room temperature with low optical power, fulfilling key criteria for deployment close to detectors.

Thus, our CV-QELM architecture consists of three stages: (i) input embedding by multimode displacement, (ii) propagation through a fixed Gaussian interferometric circuit that provides random mixing and entanglement via cascading $\hat{C}_X(s)$-gates, and (iii) readout through Gaussian compatible detection to yield the feature vector $\Phi_\mathrm{(hom)}(\mathbf{x})$. This design balances expressivity and trainability while maintaining fixed optical latency, making continuous-variable photonics a practical platform for scalable quantum extreme learning at the front-end of high-energy physics data acquisition chains. In a practical collider trigger environment, the classical detector features must be encoded into the optical device on an event-by-event basis. This can be achieved using electro-optic modulators that map electronic signals into optical phase or amplitude shifts, thereby implementing the displacement operations $\hat{D}(\alpha(\mathbf{x}))$. Modern integrated modulators based on silicon photonics typically achieve bandwidths in the tens of GHz range, corresponding to response times of order 10–100\,ps. This is substantially faster than the 25\,ns bunch-crossing interval of the LHC, and therefore compatible with event-by-event encoding at the 40\,MHz collision rate.


\begin{figure}[t!]
    \centering
    \begin{subfigure}{0.5\textwidth}
        \centering
        \includegraphics[width=\textwidth]{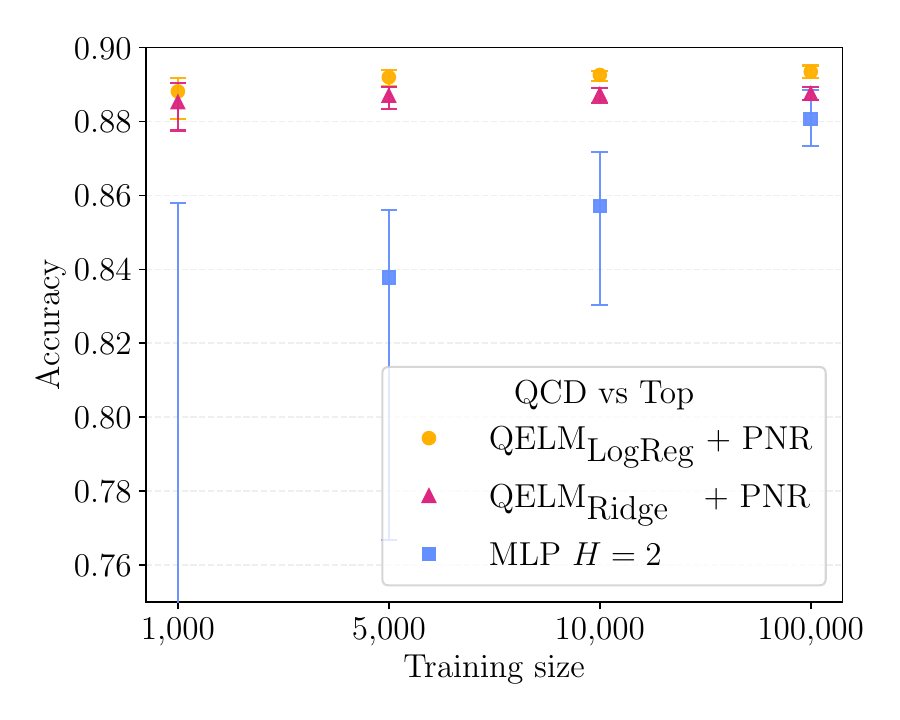}
    \end{subfigure}%
    \hfill
    \begin{subfigure}{0.5\textwidth}
        \centering
        \includegraphics[width=\textwidth]{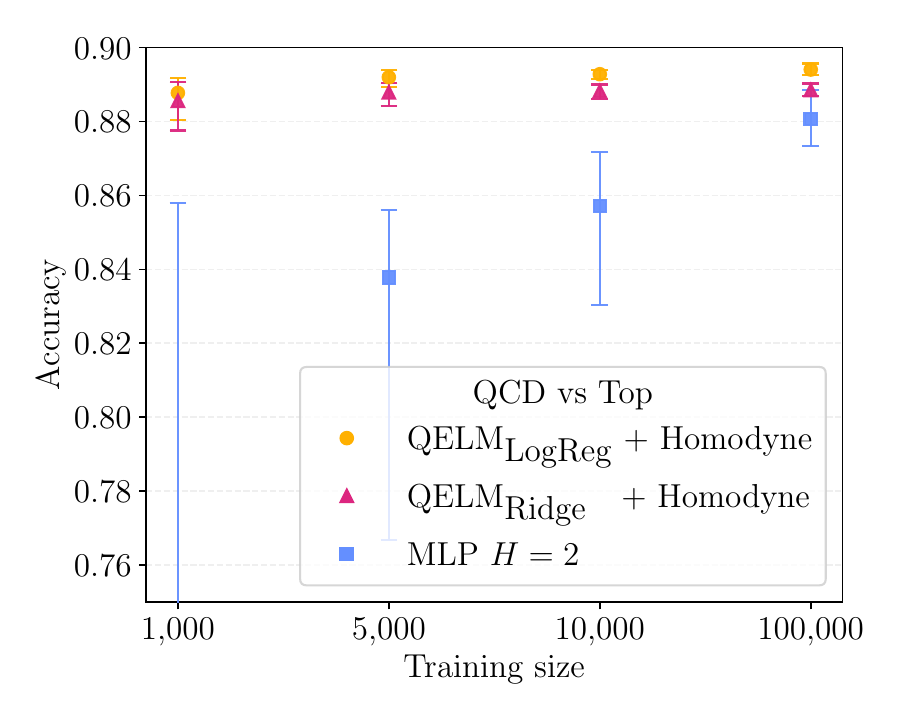}
    \end{subfigure}
    \caption{Scaling of test accuracy with training set size for the QCD versus Top jet classification task with $F=16$ features~\cite{pierini_2020_3602260}. Mean test accuracy is shown for the CV-QELM with (left) PNR and (right) homodyne measurements using both logistic and ridge regression readouts, compared against an MLP baseline with $H=2$. Error bars denote the standard deviation across repeated runs. The CV-QELM consistently outperforms the MLP across all training sizes, maintaining high accuracy and low variance even in the low-statistics regime.}
    \label{fig:qcdvstop_scaling}
\end{figure}

\section{Results}
\label{sec:results}

In this section we evaluate the performance of the continuous-variable quantum extreme learning machine (CV-QELM) on representative classification tasks from collider physics. We first consider top jet tagging, a benchmark problem that distinguishes boosted hadronic top quarks from QCD background jets, followed by Higgs boson identification. These tasks span different levels of feature complexity and class separability, allowing us to assess the robustness and data efficiency of the QELM relative to compact classical baselines.

Let $F$ denote the number of input features, $H$ the number of nodes in the hidden layer of the MLP baseline, and $M$ the number of optical modes in the quantum substrate\footnote{Note, for the studies conducted in this paper, we will use $F=M$ and embed a single feature on each qumode. In general, one could use a substrate with more or fewer optical modes than $F$.}. As outlined in Section~\ref{sec:cvqelm}, we consider two measurement schemes. In the first, PNR detectors measure the observable vector $\hat{\mathbf{O}}(\mathbf{x})$ defined in Equation~\ref{eqn:pnrMeas}, yielding a feature vector $\Phi(\mathbf{x})$ of dimension $R = 2M$. In the second, homodyne measurements evaluate the observable vector $\hat{\mathbf{O}}_\mathrm{hom}(\mathbf{x})$ from Equation~\ref{eqn:homMeas}, producing a feature vector $\Phi_\mathrm{hom}(\mathbf{x})$ of dimension $R = 5M$. In both cases, $\Phi_{(\mathrm{hom})}(\mathbf{x}) \in \mathbb{R}^{R}$ denotes the resulting random feature vector which is then used downstream for classification. For binary classification, we use a single output logit, so we write $C_{\mathrm{out}} = 1$.

The CV-QELM trains only a linear readout on top of the concatenation of the fixed random features $\Phi_{(\mathrm{hom})}(\mathbf{x})$ and the input features $\mathbf{x}$ of dimension $F$. This design allows the readout to preserve discriminative information directly present in the original inputs while the quantum substrate supplies richer non-linear feature combinations, and we find that combining both improves model accuracy and training stability. With one logit, the number of trainable parameters is therefore
\begin{equation}
n_{\mathrm{train}}^{\mathrm{QELM}} \;=\; F+R + 1 \qquad \text{(weights plus bias)}~.
\end{equation}
The MLP baseline, trained on the classical features only, has one hidden layer of width $H$ with a LeakyReLU activation function. With $F$ inputs and one logit, the number of trainable parameters is
\begin{align}\label{eqn:mlpFeats}
n_{\mathrm{train}}^{\mathrm{MLP}}(H)
\;&=\; \underbrace{F H}_{\text{input}\to\text{hidden}} \,+\, \underbrace{H}_{\text{hidden biases}} \,+\, \underbrace{H}_{\text{hidden}\to\text{output}} \,+\, \underbrace{1}_{\text{output bias}} \nonumber\\[1em]
&=\; H\,(F+2) + 1~.
\end{align}
In the following studies, $H$ is chosen to test the CV-QELM's performance against MLPs with comparable and larger numbers of trainable parameters.


\subsection{Top jet tagging: QCD vs Top}
\label{sec:topjet}

\begin{figure}[t!]
    \centering
    \begin{subfigure}{0.5\textwidth}
        \centering
        \includegraphics[width=\textwidth]{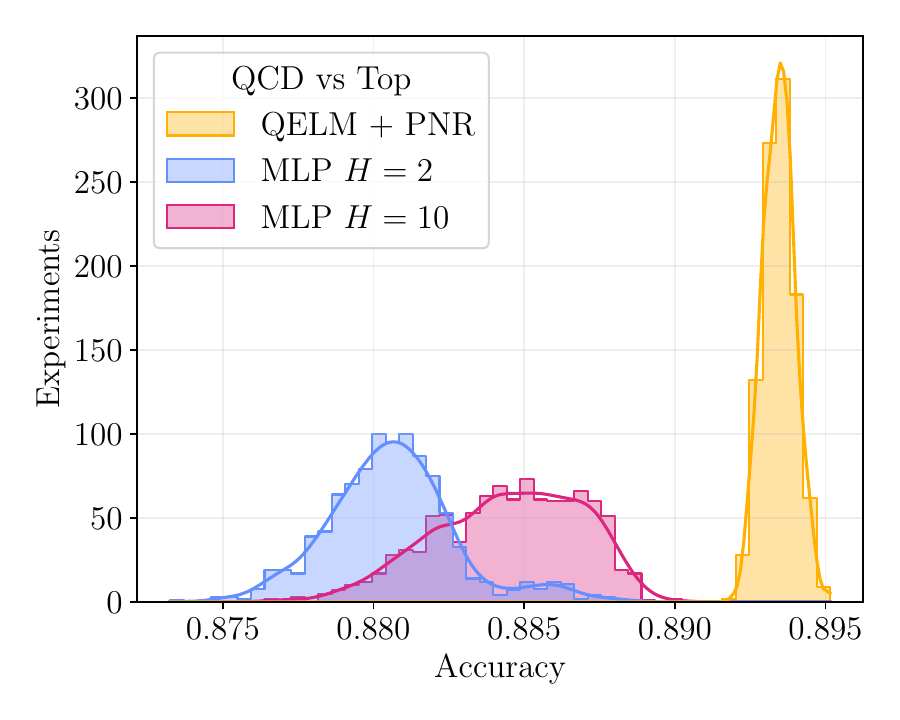}
    \end{subfigure}%
    \hfill
    \begin{subfigure}{0.5\textwidth}
        \centering
        \includegraphics[width=\textwidth]{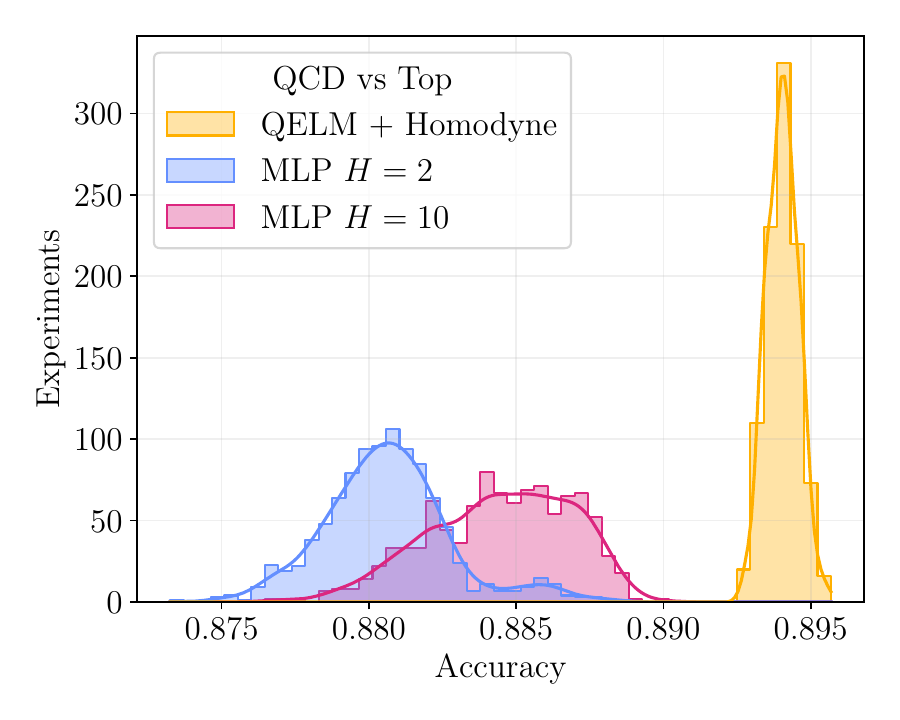}
    \end{subfigure}
    \caption{QCD versus Top jet classification with $F=16$ features and $n_{\mathrm{samples}}=10^5$ from the \textsc{hls4ml} dataset~\cite{pierini_2020_3602260} using logistic regression on the CV-QELM outputs. Accuracy distributions are shown for the CV-QELM with (left) PNR and (right) homodyne measurements, compared against MLP baselines with $H=2$ and $H=10$. The CV-QELM achieves higher mean accuracy and substantially lower variance than both MLP baselines, demonstrating the stability and expressiveness of the Gaussian photonic random feature representation.}
    \label{fig:qcdvstop_feats16}
\end{figure}

We benchmark the CV-QELM on a standard collider classification task, distinguishing top jets from QCD background jets. The study uses the public \textsc{hls4ml} LHC jet dataset~\cite{pierini_2020_3602260}, in which each jet is represented by $F=16$ engineered high-level observables that encapsulating the jet's substructure. To provide classical baselines, we train single-hidden-layer MLPs with widths $H \in \{2,10\}$. From Equation~\ref{eqn:mlpFeats}, these values correspond respectively to parameter counts comparable to and exceeding those of the CV-QELM readout.

For the CV-QELM, as outlined in Section~\ref{sec:cvqelm}, the input features are embedded as quadrature displacements on the qumode device, such that $M=16$. They are then propagated through a fixed randomly drawn Gaussian circuit, $\hat U_G$, and measured using either PNR or homodyne detection to produce $\Phi_\mathrm{(hom)}(\mathbf{x})$. The feature vector is then classified using either logistic regression or ridge regression. The MLP baselines are trained with cross-entropy loss, early stopping, and a small grid search over weight decay. Identical training, validation, and test splits are used across models. 

To assess data efficiency, we vary the number of training examples $n_{\mathrm{samples}}\in\{10^3,\,5\times10^3,\,10^4,\,10^5\}$ and report test metrics averaged over independent initialisations. Figure~\ref{fig:qcdvstop_scaling} shows the scaling of test accuracy with training set size for both measurement schemes (PNR and homodyne) and both regression methods (logistic and ridge), alongside an MLP baseline with a single hidden layer of width $H=2$.  

Across all training sizes, the CV-QELM consistently outperforms the classical MLP baseline. The CV-QELM also exhibits remarkable stability, maintaining high and nearly constant accuracy with minimal variance across independent runs. Logistic and ridge regression yield comparable performance, with logistic regression marginally higher on average. This demonstrates that ridge regression provides competitive accuracy while retaining its analytical simplicity and low training cost. At larger training sizes, the homodyne measurement scheme slightly outperforms the PNR-based variant, indicating that homodyne detection combined with ridge regression offers an especially fast and efficient low-latency classification routine.

\begin{figure}[t!]
    \centering
    \begin{subfigure}{0.5\textwidth}
        \centering
        \includegraphics[width=\textwidth]{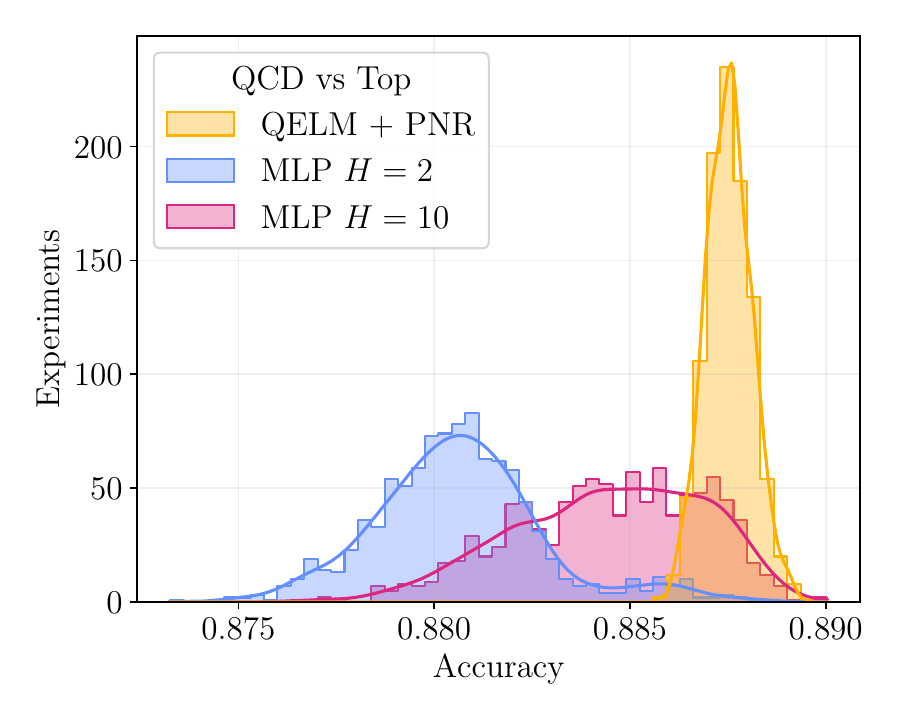}
    \end{subfigure}%
    \hfill
    \begin{subfigure}{0.5\textwidth}
        \centering
        \includegraphics[width=\textwidth]{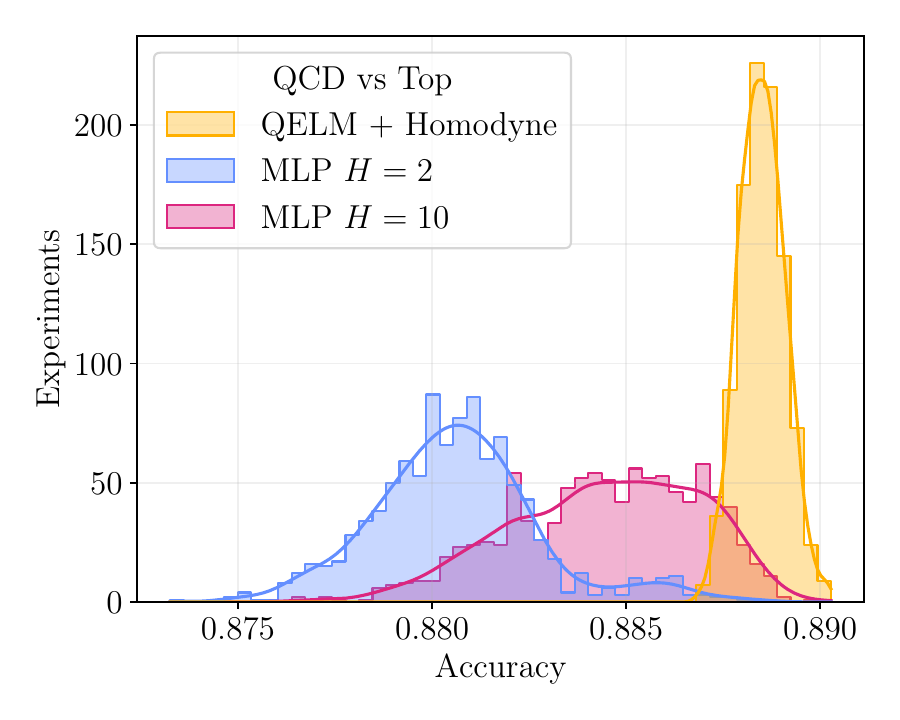}
    \end{subfigure}
    \caption{QCD versus Top jet classification with $F=16$ features and $n_{\mathrm{samples}}=10^5$ from the \textsc{hls4ml} dataset~\cite{pierini_2020_3602260} using ridge regression on the CV-QELM outputs. Accuracy distributions are shown for the CV-QELM with (left) PNR and (right) homodyne measurements, compared against MLP baselines with $H=2$ and $H=10$. The CV-QELM achieves higher mean accuracy and markedly lower variance than both MLP baselines, highlighting the robustness and efficiency of the analytic ridge regression readout in combination with Gaussian photonic random features.}
    \label{fig:qcdvstop_feats16_ridge}
\end{figure}

Figures~\ref{fig:qcdvstop_feats16} and~\ref{fig:qcdvstop_feats16_ridge} further illustrate this performance gain for $n_{\mathrm{samples}} = 10^5$, showing logistic and ridge regression on the CV-QELM outputs, respectively, compared with MLP baselines of $H=2$ and $H=10$. The CV-QELM surpasses both baselines, including the wider $H=10$ network, which from Equation~\ref{eqn:mlpFeats} contains $181$ trainable parameters. The accuracy distributions demonstrate the robustness of the CV-QELM, which exhibits substantially smaller variance across repeated runs than the MLPs due to its fixed feature map and analytic readout fit. This combination of accuracy, predictable training cost, and fixed, nanosecond-scale inference latency supports the use of CV-QELMs as low-overhead front-end classifiers in collider data acquisition.

\subsection{Higgs boson identification}
\label{sec:higgs}

\begin{figure}[t!]
    \centering
    \begin{subfigure}{0.5\textwidth}
        \centering
        \includegraphics[width=\textwidth]{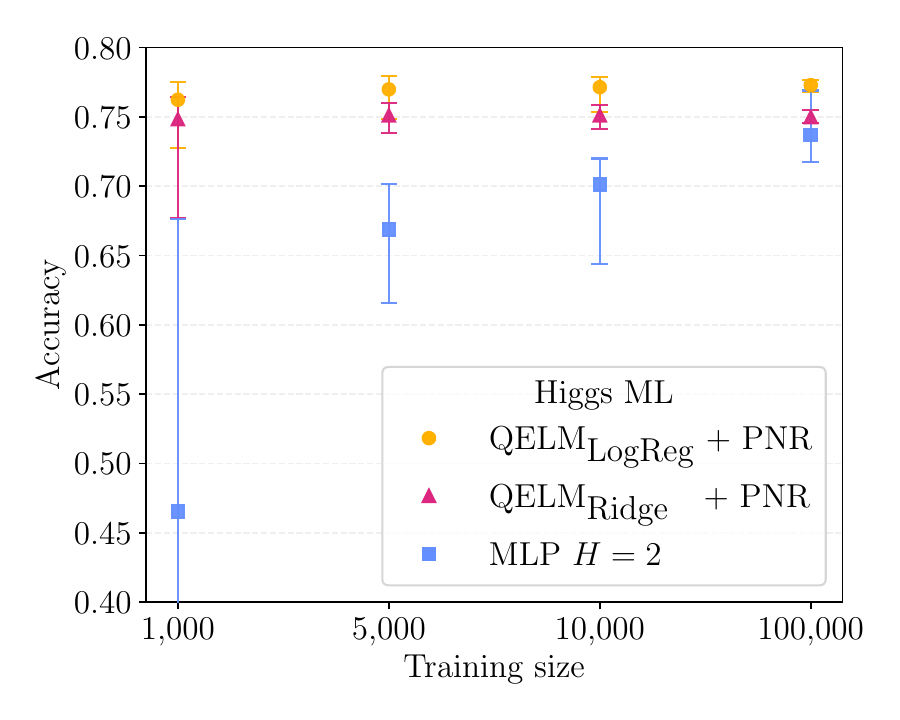}
    \end{subfigure}%
    \hfill
    \begin{subfigure}{0.5\textwidth}
        \centering
        \includegraphics[width=\textwidth]{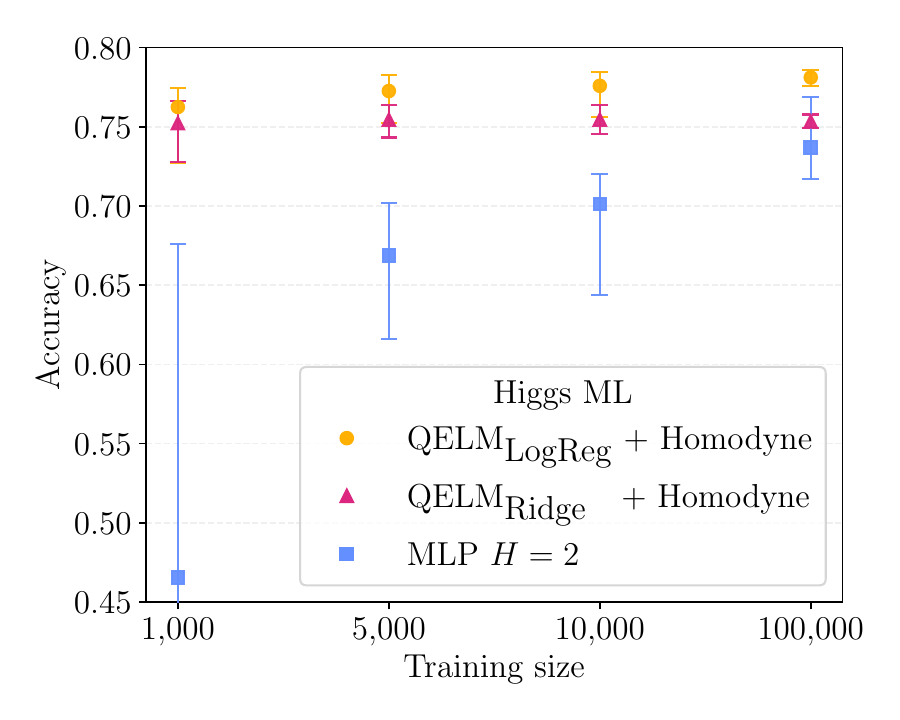}
    \end{subfigure}
    \caption{Scaling with training set size on the $F=10$ Higgs task~\cite{higgs-boson}. Mean test accuracy versus $n_{\mathrm{samples}}$ for the CV QELM with (left) PNR and (right) Homodyne measurements and the MLP with $H=2$. Shaded bands show the standard deviation across initialisations. The QELM maintains a consistent advantage from $10^3$ to $10^5$ examples, mirroring the trends observed in Sec.~\ref{sec:topjet}.}
    \label{fig:higgs_scaling}
\end{figure}

We evaluate the continuous-variable QELM on the \textsc{Higgs Boson Machine Learning Challenge} dataset~\cite{higgs-boson}, a binary classification task distinguishing simulated Higgs signal from background. We use the curated subset of $F=10$ engineered features to enable direct comparison with MLP baselines and to mirror the setup in Sec.~\ref{sec:topjet}. Following the notation introduced above, the CV-QELM, which is constructed from $M=10$ qumodes, produces a random feature vector $\Phi_\mathrm{(hom)}(\mathbf{x})\in\mathbb{R}^{R}$ and trains a single linear readout with $n_{\mathrm{train}}^{\mathrm{QELM}} = F + R + 1$, while a single-hidden-layer MLP with width $H$ has $n_{\mathrm{train}}^{\mathrm{MLP}}(H) = H(F + 2) + 1$. We again consider $H\in\{2,10\}$ to provide a parameter-matched small baseline and a moderately larger one. Training, validation, and test splits are identical across models. 

The experiments are carried out in the same way as for the top tagging benchmark varying the number of training examples $n_{\mathrm{samples}}\in\{10^3, 5\times10^3, 10^4, 10^5\}$ and quantify performance by reporting the mean test accuracy averaged over independent runs. Figure~\ref{fig:higgs_scaling} shows the scaling of test accuracy with training set size for both measurement schemes (PNR and homodyne) and both regression methods (logistic and ridge), alongside an MLP baseline with a single hidden layer of width $H=2$. The CV-QELM consistently outperforms the parameter-matched MLP across all training sizes, with the margin most pronounced in the low-data regime, reflecting the stability of the analytic readout fit and the absence of stochastic optimisation variance. Absolute accuracies are lower than in the QCD versus Top task, consistent with the weaker class separability of the Higgs dataset; however, the relative advantage of the QELM persists. Training time for the QELM remains dominated by a single linear solve that scales predictably with $n_{\mathrm{samples}}$, whereas the MLP requires iterative optimisation and hyperparameter tuning.

\begin{figure}[t!]
\centering
\begin{subfigure}{0.5\textwidth}
    \centering
    \includegraphics[width=\textwidth]{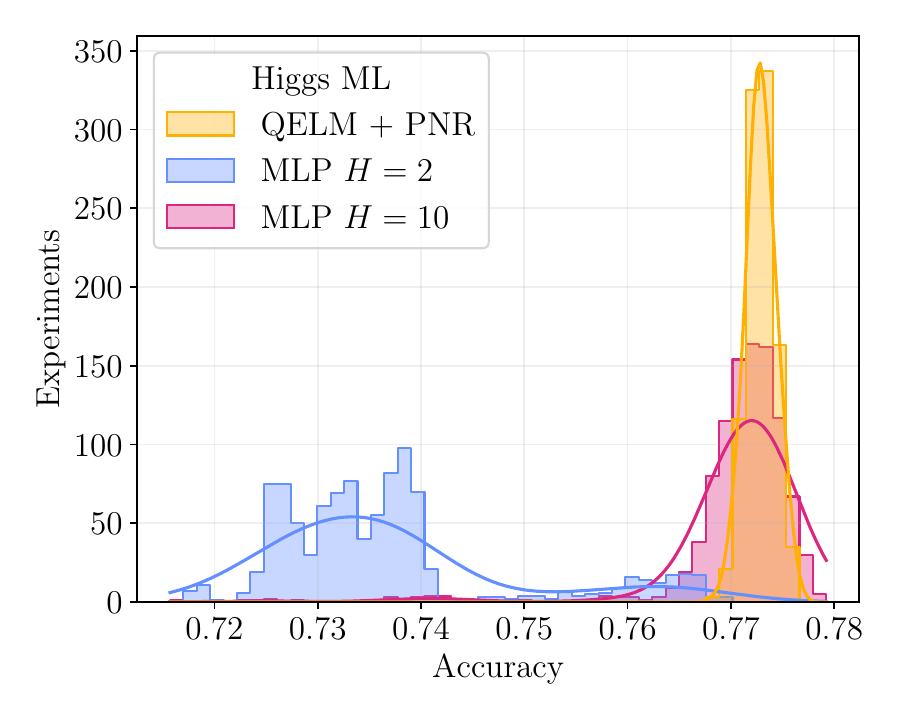}
\end{subfigure}%
\hfill
\begin{subfigure}{0.5\textwidth}
    \centering
    \includegraphics[width=\textwidth]{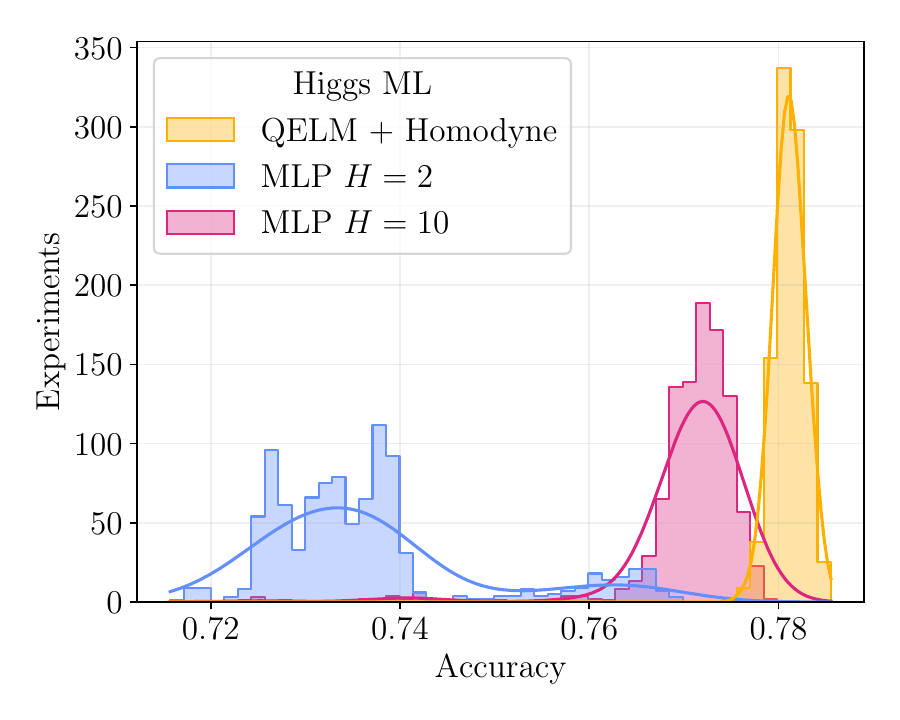}
\end{subfigure}
\caption{Higgs versus background classification with $F=10$ features and $n_{\mathrm{samples}}=10^5$ of the \textsc{Higgs ML} dataset~\cite{higgs-boson} using logistic regression on the CV-QELM outputs. Accuracy distributions are shown for the CV-QELM with (left) PNR and (right) homodyne measurements, compared against MLP baselines with $H=2$ and $H=10$. The CV-QELM matches the wider $H=10$ MLP under PNR detection and slightly surpasses it with homodyne detection, while maintaining substantially lower variance across runs.}
    \label{fig:higgs_n1e5}
\end{figure}

\begin{figure}[t!]
\centering
\begin{subfigure}{0.5\textwidth}
    \centering
    \includegraphics[width=\textwidth]{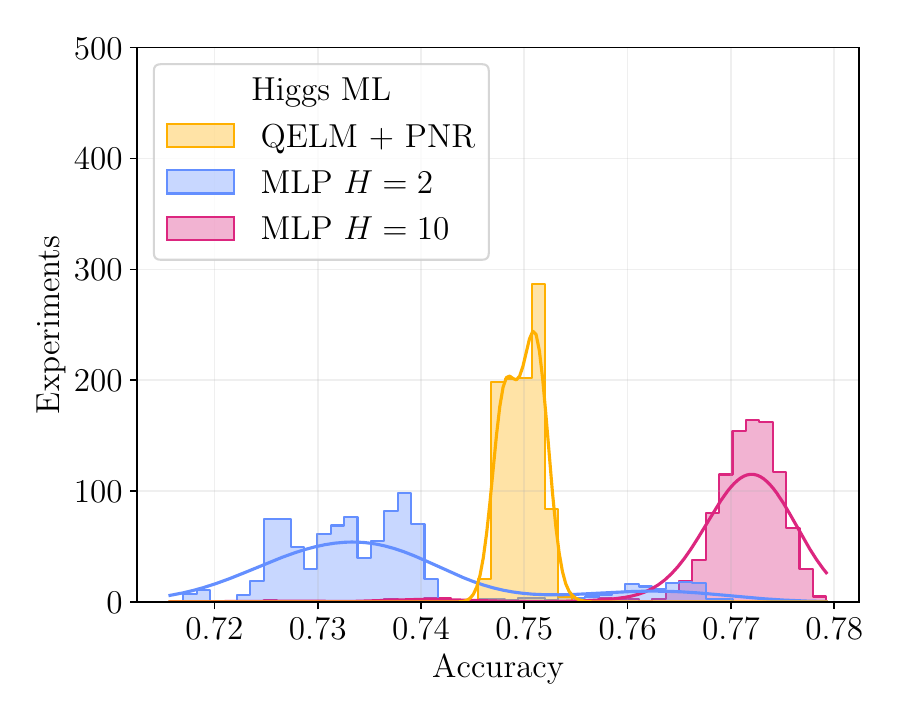}
\end{subfigure}%
\hfill
\begin{subfigure}{0.5\textwidth}
    \centering
    \includegraphics[width=\textwidth]{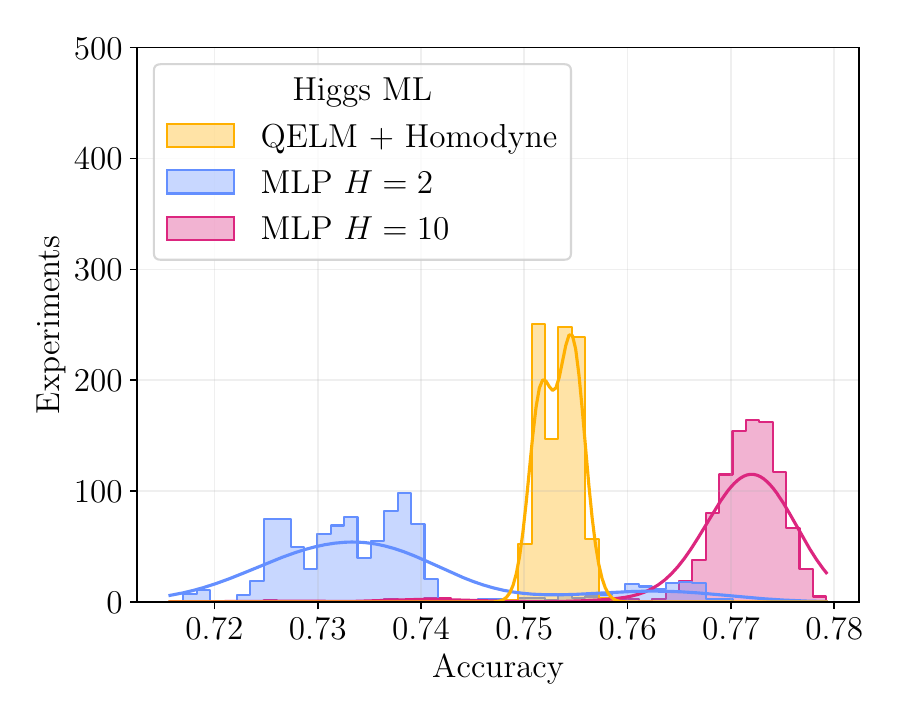}
\end{subfigure}
\caption{Higgs versus background classification with $F=10$ features and $n_{\mathrm{samples}}=10^5$ of the \textsc{Higgs ML} dataset~\cite{higgs-boson} using ridge regression on the CV-QELM outputs. Accuracy distributions are shown for the CV-QELM with (left) PNR and (right) homodyne measurements, compared against MLP baselines with $H=2$ and $H=10$. Both QELM variants outperform the parameter-matched $H=2$ MLP but fall slightly short of the $H=10$ model, while maintaining far lower run-to-run variance due to the deterministic analytic readout.}
    \label{fig:higgs_n1e5_ridge}
\end{figure}

To further probe the QELM framework, we compare its performance against an MLP with a wider hidden layer of width $H=10$, which from Equation~\ref{eqn:mlpFeats} has $121$ trainable parameters. Figures~\ref{fig:higgs_n1e5} and~\ref{fig:higgs_n1e5_ridge} show results for a training size of $n_{\mathrm{samples}}=10^5$ using logistic and ridge regression on the CV-QELM outputs, respectively, compared against MLP baselines with $H=2$ and $H=10$. In the logistic case, the CV-QELM achieves comparable performance to the $H=10$ MLP under PNR detection and slightly surpasses it under homodyne detection. For ridge regression, both measurement schemes yield accuracies marginally below the $H=10$ MLP but consistently above the $H=2$ baseline. In all cases, the CV-QELM exhibits markedly smaller variance across runs, reflecting the reproducibility of the analytic readout training and the stability of the photonic random feature map.


\section{Summary and Conclusions}
\label{sec:summary}

We investigated continuous-variable quantum extreme learning machines (CV-QELMs) as fast, low-overhead front-ends for collider data processing. The architecture employs displacement based input embedding across $M$ optical modes, a fixed-time Gaussian photonic quantum substrate, $\hat{U}_G$, that mixes and entangles the modes, and photon-number-resolving (PNR) or homodyne readout to generate a random feature vector, $\Phi_\mathrm{(hom)}(\mathbf{x})$, of dimension $R$. Only a linear readout layer is trained via a single regularised solve, making retraining extremely fast. In this work, we considered both logistic regression and ridge regression as the linear models used to train the output of the quantum substrate. The optical path and detector response set both the analytic and inference latency. The design emphasises predictable, non-stochastic behaviour, straightforward reconfigurability, and low-power operation, all of which are relevant for future Level-1 trigger integration or on-detector data reduction.

We established a parameter-matched evaluation protocol against classical baselines. For a multilayer perceptron (MLP) with one hidden layer of width $H$ and $F$ input features, the number of trainable parameters is $H(F + 2) + 1$. For the CV-QELM, only the linear readout parameters, $R + 1$, are trained. Using identical data splits and averaging over independent runs, we benchmarked two collider-relevant tasks: top jet tagging and Higgs boson identification.

For the \textsc{hls4ml} LHC jet dataset with $F=16$ engineered features, the CV-QELM consistently outperformed the $H=2$ MLP across all training sizes, with the largest margin in the low-statistics regime. At $n_{\mathrm{samples}} = 10^5$, the CV-QELM also surpasses the wider $H=10$ MLP, demonstrating that the Gaussian photonic random feature map provides a compact and expressive representation. For the \textsc{Higgs Boson Machine Learning Challenge} dataset with $F=10$ features, the CV-QELM again exceeded the $H=2$ baseline across all sample sizes. With logistic regression, it achieved accuracy comparable to the $H=10$ MLP under PNR detection and slightly higher under homodyne detection, while ridge regression yielded results marginally below the $H=10$ model but above the $H=2$ baseline. In all cases, the CV-QELM displayed significantly lower variance across runs, reflecting the reproducibility of the analytic readout training and the stability of the photonic feature map.

Beyond accuracy, we quantified the practical advantages that follow from analytic regression based training. Since only a linear readout is fitted through a single regularised regression, training scales predictably with sample size and avoids optimisation variance, learning rate tuning, or early stopping heuristics. Inference latency is fixed and extremely low, on the order of nanoseconds for homodyne detection, and independent of event complexity within the designed optical depth. Combined with the low optical power and room-temperature operation of integrated photonics, these properties support deployment close to the detector.

In summary, we demonstrated that a fixed Gaussian photonic quantum substrate coupled to an analytic readout can deliver competitive or superior classification performance to classical networks with similar or larger numbers of trainable parameters, while offering deterministic, low-latency inference and negligible training overhead. These results establish CV-QELMs as a promising building block for real-time pattern recognition in future collider experiments. A natural next step is hardware co-design and on-bench validation of latency, power, and noise characteristics, including systematic sweeps over substrate depth, squeezing strength, and readout to extract design scaling laws. The approach can also be extended to multi-class tagging, calibration-sensitive regression, and robustness studies under domain shift.

\paragraph{Acknowledgement:} The authors acknowledge the support of Schmidt Sciences and IPPP Durham.
\paragraph{Author Contributions:} M.S. conceived of the initial idea. B.M. and S.W. designed and conducted the experiments. M.S. and S.W. wrote the manuscript. All authors reviewed the manuscript. 
\paragraph{Declaration of Competing Interests:} All authors declare that there are no competing interests in this publication.

\bibliographystyle{inspire}
\bibliography{bibliography} 

\end{document}